\documentclass[onecolumn]{revtex4}
\usepackage{amsmath,amssymb,graphics,epsfig,subfigure}
\usepackage{color}
\usepackage[colorlinks,linkcolor=red,anchorcolor=red,citecolor=green]{hyperref}
\usepackage{setspace}
\usepackage{booktabs}
\usepackage{float}
\usepackage[scheme=plain]{ctex}

\begin{document}
\thispagestyle{empty}

\title{Mean first passage time and the Kramers escape rate of phase transitions for the Bardeen-AdS-class black hole}

\author{Chen Ma$^{1,2}$, Bin Wu$^{1,3,4,5}$\footnote{E-mail: binwu@nwu.edu.cn(Corresponding author)}, and Zhen-Ming Xu$^{1,3,4,5}$}

\affiliation{$^{1}$School of Physics, Northwest University, Xi'an 710127, China\\
$^{2}$Institute of Modern Physics, Northwest University, Xi'an 710127, China\\
$^{3}$Peng Huanwu Center for Fundamental Theory, Xi'an 710127, China\\
$^{4}$Shaanxi Key Laboratory for Theoretical Physics Frontiers, Xi'an 710127, China\\
$^{5}$Fundamental Discipline Research Center for Quantum Science and technology of Shaanxi Province, Xi'an 710127, China}

\begin{abstract}
In this study, by utilizing the constructed generalized free energy alongside the Mean First-Passage Time and the Kramers escape rate from stochastic dynamics, we have obtained a comprehensive landscape of the phase transitions for the Bardeen-AdS-class black hole. This black hole model admits two distinct categories of solutions. Type I black holes feature a regular black hole solution, and Type II black holes possess a vacuum state solution. In the phase transition between the small black hole and the large black hole for Type I, the process may pass through a stable, metastable, or unstable regular black hole as an intermediate state. In contrast, for Type II black holes, the phase transition occurs exclusively between the vacuum state and the small black hole, and the transition process does not involve any regular black hole intermediate states.
\end{abstract}

\maketitle

\vspace{0.3em}

\section{Introduction}

The center of a general black hole solution harbors a gravitational singularity, a point where spacetime curvature diverges and all known physical laws cease to be valid. This constitutes a major theoretical flaw, and quantum theory also prohibits the appearance of infinite curvature. However, the singularity theorems proposed by Hawking and Penrose predict that, under certain conditions, a black hole formed from stellar collapse must possess a singularity\cite{hawking1973large}. Nevertheless, we can potentially avert the singularity by introducing a coupled matter field\cite{ayon2000bardeen,liu2020quasi,cisterna2020quasitopological,li2024regular,bronnikov2005regular}. One such example is the Bardeen black hole, which is obtained by coupling Einstein gravity to a nonlinear electromagnetic field. Such black holes without a singularity are termed regular black holes. Furthermore, regular black holes can also be obtained through certain modified theories of gravity\cite{nicolini2006noncommutative,eichhorn2023black,ashtekar2023regular,nicolini2023strings,bueno2025regular,konoplya2024dymnikova,ghosh2023regularized}.

Ever since the establishment of black hole thermodynamics, there has been a natural curiosity regarding the thermodynamics of regular black holes, leading to analyses of their thermodynamic behavior~\cite{rasheed1997non,born1934foundations,gunasekaran2012extended,ma2014corrected,lan2021entropy,lan2023entropy}. However, several issues arose at this point. For instance, the requirements of the First Law led to a conflict between the entropy-area law and the Hawking temperature~\cite{ma2014corrected,lan2021entropy}. Simultaneously, the electromagnetic potential also required redefinition~\cite{lan2023entropy}. This conflict primarily stems from the fact that the energy and magnetic charge of regular black holes are generally dependent on the coupling coefficients of the matter field. Consequently, their relevant integration constants cannot deviate from the coupling constants. Crucially, in thermodynamics, the variation of energy must be considered simultaneously with the variation of other coupling constants. This requirement ultimately leads to an inconsistency between the thermodynamic quantities and the First Law.

In the literature~\cite{wu2025thermodynamics}, the authors proposed a solution where they treat the integration constants, such as energy and charge, not as coupling parameters, thereby exploring a more generalized black hole solution~\cite{rasheed1997non,ma2014corrected,zhang2018first,guo2024recovery}. In these solutions, regular black holes exist only when the energy and charge adopt specific values.
Therefore, our central question is: Is the regular black hole characterized as a stable (or metastable) intermediate or an unstable transition state during the phase transition process?
We aim to explore this question using non-equilibrium statistical methods, hoping to gain a deeper understanding of the formation mechanism of regular black holes.

Given that black hole phase transitions are not necessarily quasi-static processes, researchers have made several attempts to study the dynamics of black hole thermodynamic phase transitions. These efforts seek to uncover the underlying structure of the phase transition by establishing a framework of relativistic stochastic statistical physics~\cite{cai2023relativistic}. Specifically, some studies attempt to obtain dynamic information about the phase transition by utilizing the mean first passage time (MFPT) within the context of stochastic motion~\cite{li2020thermodynamics,wei2021observing,cai2021oscillatory,yue2025transit,li2024thermodynamics}. Other works focus on employing generalized free energy to determine the rate behavior of the phase transition~\cite{xu2023rate,ma2025kramers,wang2024thermodynamic,xu2024thermodynamic,du2024topology,afshar2025kramer,sadeghi2025phase}.

In this paper, Our discussion centers on the case of a Bardeen-AdS-class black hole, simultaneously utilizing the MFPT and the Kramers escape rate to analyze the black hole phase transition. The paper is organized as follows. In Sec.~\ref{2}, we present the metric and thermodynamics of the Bardeen-AdS-class black holes. In Sec.~\ref{3}, the definition of the generalized free energy is established, and the MFPT and the Kramers escape rate are derived via the Fokker-Planck equation. These theoretical frameworks are then employed in Sec.~\ref{4} to analyze the phase transition behaviors of the Bardeen-AdS-class black holes. Finally, Sec.~\ref{5} is devoted to the discussion of our results.

\section{The Bardeen-AdS-class black hole}
\label{2}
Considering the coupling of Einstein's gravity and nonlinear electromagnetic fields, the Bardeen(-AdS) black hole solution  can be obtained. To circumvent the ``coupling constant issues'' where the mass and charge of regular black holes are intrinsically linked to the matter field's coupling parameters, the authors introduced a modified non-linear electromagnetic Lagrangian~\cite{wu2025thermodynamics}
\begin{equation}
\mathcal{L}_m=\frac{6 m_0}{q_0^3}\left(\frac{\sqrt{q_0^2 F_{\mu \nu} F^{\mu \nu} / 2}}{1+\sqrt{q_0^2 F_{\mu \nu} F^{\mu \nu} / 2}}\right)^{5 / 2}.
\end{equation}
By minimally coupling of this Lagrangian and the Einstein's gravity, the action is given by
\begin{equation}
I=\frac{1}{16 \pi} \int d^4 x \sqrt{-g}\left(R+\frac{6}{L^2}-\frac{6 m_0}{q_0^3}\left(\frac{\sqrt{q_0^2 F_{\mu \nu} F^{\mu \nu} / 2}}{1+\sqrt{q_0^2 F_{\mu \nu} F^{\mu \nu} / 2}}\right)^{5 / 2}\right).\label{act1}
\end{equation}

In the original Bardeen-AdS black hole solution, the parameters $m_0$ and $q_0$ possess a dual identity. On one hand, they function as coupling constants of the interaction; on the other hand, they represent the black hole's mass and magnetic charge. This implies that altering the black hole's mass or magnetic charge would effectively change the entire underlying theory, which naturally gives rise to the various thermodynamic inconsistencies mentioned previously. To resolve this issue, it is necessary to decouple these roles. Here the parameters $m_0$ and $q_0$ are fixed as constants purely associated with the coupling, while the actual black hole parameters are redefined in terms of $m$ and $q_m$. Solving the field equation corresponding to the action (\ref{act1}), the Bardeen-AdS-class black hole solution can be obtained
\begin{equation}
f(r)=\frac{r^2}{L^2}+1-\frac{m}{r}+m_0\left(\frac{q_m}{q_0}\right)^{3 / 2}\left(\frac{1}{r}-\frac{r^2}{\left(r^2+q_m q_0\right)^{3 / 2}}\right),\label{bh}
\end{equation}
and
\begin{equation}
F=q_m \sin \theta d \theta \wedge d \phi.
\end{equation}
The energy and magnetic charge of black hole are given by
\begin{equation}
M=\frac{1}{2} m, \quad Q_m=q_m.
\end{equation}
When the black hole parameters are equal to the model parameters, i.e., $m=m_0$, $q_m=q_0$, the Bardeen-AdS-class black hole reduces to the Bardeen-AdS black hole.

The event horizon of the black hole is located at the largest root of $f\left(r_+\right)=0$, then the Hawking temperature in terms of event horizon radius $r_+$ is given by
\begin{equation}
T_h=\left.\frac{1}{4 \pi} \partial_r f\left(r, m\left(r_+\right)\right)\right|_{r=r_+}=\frac{1}{4 \pi}\left(\frac{1}{r_{+}}+3 \frac{r_{+}}{L^2}-3 m_0 r_{+} \sqrt{\frac{q_m^5}{q_0\left(q_0 q_m+r_{+}^2\right)^5}}\right).
\end{equation}
It should be noted that, since this black hole may possess multiple horizons, $r_+$ cannot always take on a continuous range of values. For example, when the curve of $f(r)$ has two extreme points, it is impossible for $r_+$ to fall within its monotonically decreasing interval. This point affects the range of the thermodynamic potential. The corresponding mass is given by	
\begin{equation}	
M=\frac{1}{2} m=\frac{1}{2}\left(\frac{r_{+}^3}{L^2}+r_{+}+m_0\left(\frac{q_m}{q_0}\right)^{3 / 2}\left(1-\frac{r_{+}^3}{\left(r_{+}^2+q_0 q_m\right)^{3 / 2}}\right)\right) .
\end{equation}
And the other thermodynamical quantities for the Bardeen-AdS-class black holes are described by~\cite{wu2025thermodynamics}
\begin{align}
S&=\pi r_{+}^2,\\
V&=\frac{4}{3} \pi r_{+}^3,\\
\Phi_m&=\frac{3 m_0}{4 q_0} \sqrt{\frac{q_m}{q_0}}\left(1-\frac{r_{+}^5}{\left(r_{+}^2+q_0 q_m\right)^{5 / 2}}\right),\\
\phi_m&=\frac{1}{2}\left(\frac{q_m}{q_0}\right)^{3 / 2}\left(1-\frac{r_{+}^3}{\left(r_{+}^2+q_0 q_m\right)^{3 / 2}}\right),\\
\phi_q&=\frac{3 m_0}{4 q_0}\left(\frac{q_m}{q_0}\right)^{3 / 2}\left(-\frac{r_{+}^5+2 q_0 q_m r_{+}^3}{\left(r_{+}^2+q_0 q_m\right)^{5 / 2}}-1\right).
\end{align}
They satisfy the extended first law of black hole thermodynamics
\begin{align}
d M=T_h dS+V dP+\Phi_m d Q_m+\phi_m d m_0+\phi_q dq_0.
\end{align}
Where $\Phi_m$, $\phi_m$, and $\phi_q$ are the conjugate potentials arising from the variations of the mass $M$ with respect to the magnetic charge $Q_m$ and the coupling parameters $m_0$ and $q_0$, respectively. However, it should be pointed out that if $Q_m$ and $q_0$ are treated as identical, the magnetic potential of the black hole would become $\Phi_m + \phi_q$, which deviates from the conventional definition.

We can derive the Smarr relation using the method of scaling analysis. Considering that the dimensions of $M$, $m_0$, $Q_m$, and $q_0$ are $[L]$, the dimension of $S$ is $[L]^2$, and the dimension of $P$ is $[L]^{-2}$, we have
	\begin{align}
	\lambda M= M(\lambda^2 S, \lambda^{-2} P, \lambda Q_m, \lambda m_0, \lambda q_0).
	\end{align}
	By differentiating with respect to $\lambda$ and setting $\lambda=1$, the Smarr relation is obtained
	\begin{align}
	M=2T_h S-2V P+\Phi_m Q_m+\phi_m m_0+\phi_q q_0.
	\end{align}
Substituting the explicit expressions of thermodynamic variables into the Smarr relation leads to its consistent verification. It follows that this thermodynamic construction, which separates the integration constants from the coupling parameters, is self-consistent.

\section{Generalized free energy, the mean first passage time and the Kramers escape rate}
\label{3}

\begin{figure}[htb]
	\centering
		\includegraphics[width=65 mm]{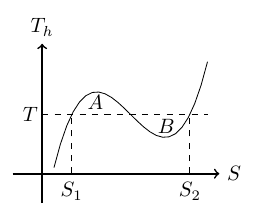}
	\caption{The Maxwell equal area law of the thermodynamic system of black holes~\cite{ma2025kramers}.}\label{fig1}	
\end{figure}

In the extended phase space of black hole thermodynamics, the phase transition of black hole can be determined by the Maxwell equal area law.  For thermodynamic system of black holes, we can obtain its equation of state, i.e., the relationship $T_h=T_h(S,P)$ between the Hawking temperature $T_h$ and the entropy $S$, and the pressure $P$ of the black hole. The Maxwell equal area law states that there exists an isotherm $T$ such that $\text{Area}_{\text{A}} = \text{Area}_{\text{B}}$, as shown in Fig~\ref{fig1}. This physical description is mathematically expressed on the $T_h-S$ plane by
\begin{align}
\int_{S_1}^{S_2}T_h dS=T\cdot(S_2-S_1) ~~\Rightarrow~~\int_{S_1}^{S_2}(T_h-T)dS=0.
\label{TS}
\end{align}

Now, based on the definition of generalized free energy in reference~\cite{xu2023rate,ma2025kramers,wang2024thermodynamic,xu2024thermodynamic}, we release the Maxwell equal area law mentioned above, and introduce generalized free energy $\mathcal{U}$ (or called the thermal potential) as
\begin{align}
\mathcal{U}=\int(T_h-T)dS,
	\label{U}
\end{align}
which are precisely the indefinite integral expressions of the above equations~\eqref{TS}. 

The central idea of a free energy landscape is to relax the constraints imposed by the equation of state, allowing internal variables to vary independently, thereby creating an energy hypersurface whose extrema correspond to equilibrium states. To clarify the notation, we use calligraphic symbols (e.g., \(\mathcal{F}, \mathcal{G}\)) for the \textit{off-shell} generalized free energy functions, which are defined for arbitrary values of internal variables and describe the system even when it is not in equilibrium. The standard \textit{on-shell} free energies, which represent true equilibrium states, are denoted by roman letters (e.g., \(F, G\)).

Since this is an indefinite integral, the result naturally includes an integration constant $C$. The choice of the integration constant $C$, which shifts the entire free energy landscape vertically, does not alter the location of the minima (stable states), and the height of the potential barriers. We typically set $C=0$. This choice is made not only for the sake of convenience, more importantly, because it ensures that the generalized free energy naturally reduces to the standard Gibbs free energy at the equilibrium state. The Hawking temperature $T_h$ represents the internal temperature of the black hole, which is a function of the thermodynamic entropy $S$. In contrast, the isotherm $T$ acts as the off-shell temperature, representing the ambient environment temperature. This $T$ can be arbitrarily assigned any positive value, independent of the black hole's state.

For generalized free energy~\eqref{U}, we can understand it is constructed under isobaric conditions, starting from the $T-S$ diagram, by relaxing the equilibrium constraint and allowing the relevant parameters to vary independently. This yields an off-shell generalized free energy (i.e., a free energy landscape), whose extremum recovers the physical equilibrium state. Based on the spirit of free energy landscape, we also refer to the thermal potential as the generalized free energy throughout this work.
\begin{equation}
	\dfrac{d \mathcal{U}}{dS}=\dfrac{d}{dS}\left(\int(T_h-T)dS\right)=0~~\Rightarrow~~T=T_h,
\end{equation}
The minima of the generalized free energy landscape correspond to thermally stable black hole states, whereas the maxima signify thermally unstable ones. Among these stable solutions, we further distinguish between metastable states (local minima) and the globally stable state (the global minimum).

Furthermore, there is another form of off-shell free energy defined as $\mathcal{G}=M-TS$~\cite{li2020thermodynamics}, where $T$ is not the system's equilibrium temperature but an arbitrary external (ensemble) temperature. This construction generalizes the Gibbs free energy $G$ by lifting the constraint $T = T_{h}$, so that the on-shell equilibrium condition $T = T_{h}$ is recovered only when $\mathcal{G}$ attains its extremum. 

For a simple thermodynamic system, the first law gives $dE = T_h dS - P dV$. Under constant thermodynamic pressure $P$, we obtain
\begin{equation}
\mathcal{U} = \int (T_h - T) dS = E + PV - TS = M - TS = \mathcal{G}.
\end{equation}
Thus, when $P$ is fixed, $\mathcal{U}$ and $\mathcal{G}$ are mathematically identical. In this sense, $\mathcal{U}$ can be viewed as a specific realization of the off-shell Gibbs free energy $\mathcal{G}$, with the integration path chosen as the deviation from equilibrium.

Regarding the Landau potential~\cite{Xu_2021}
\begin{equation}
\mathcal{L} = \int F(X, T, P) \, dX,
\end{equation}
where $X$ is an auxiliary variable interpreted as the non-equilibrium thermodynamic volume, and $F(X,T,P) = P - f(X,T)$ with $P = f(V,T)$ the equation of state. The extremum of $\mathcal{L}$ occurs when $X$ satisfies the equation of state, i.e., at the equilibrium volume $X = V_{\text{eq}}$.

The three off-shell potentials differ in which variable is taken off-shell and which thermodynamic plane they naturally belong to. $\mathcal{U}$ (on the $T$-$S$ plane) takes the ambient temperature as the off-shell variable; $\mathcal{G}$ (on the $G$-$T$ or $G$-$P$ plane) generalizes the Gibbs free energy by lifting the constraint $T = T_h$; $\mathcal{L}$ (on the $P$-$V$ plane) uses the non‑equilibrium volume $X$ as the off-shell variable.

Despite these differences, all three share the same physical essence: they construct a free energy landscape whose extremum recovers the true equilibrium state. The choice is therefore a matter of convenience, dictated by which thermodynamic plane best illustrates the process under study. In this work we adopt $\mathcal{U}$ because our analysis focuses on the $T$-$S$ plane; for $G$-$T$/$G$-$P$ or $P$-$V$ analyses, $\mathcal{G}$ or $\mathcal{L}$ would be more natural. The physical conclusions are equivalent regardless of the choice.

Due to thermodynamic fluctuations, black holes exhibit distinct phase transition behaviors within this generalized free energy framework, which can be regarded as a stochastic process.
The temporal evolution and probability distribution of these black hole states (encompassing both on-shell and off-shell states) can be described by the Fokker-Planck equation.
If the black hole temperature is significantly lower than the potential barrier height (e.g., in a double-well potential), the probability of the state residing at the lowest point of one potential well largely exceeds the probability of it reaching the top of the barrier.
Even if the state does surmount the barrier peak, it will typically fall symmetrically to either side.
However, if the state settles into the lowest point of one potential well, thermodynamic fluctuations imply that, after a period of residence, there is a finite probability for it to cross the potential barrier and eventually reach the minimum point of the other potential well.

Since we focus on the transitions between states rather than the perturbations within a specific state, we start with the overdamped Fokker-Planck equation~\cite{risken1989fokker,zwanzig2001nonequilibrium}
\begin{equation}
\frac{\partial \rho(r, t)}{\partial t}=D \frac{\partial}{\partial r}\left(e^{-\beta \mathcal{U}(r)} \frac{\partial}{\partial r}\left(e^{\beta \mathcal{U}(r)} \rho(r, t)\right)\right),\label{fp}
\end{equation}
where the $\rho(r, t)$ is the probability distribution of black hole states, $\beta=1/T$ denotes the inverse temperature, $\mathcal{U}(r)$ is the generalized free energy in our work, and $D$ is the diffusion coefficient. The radius of the black hole event horizon $r$ is regarded as a order parameter of the system.
To render the Fokker-Planck equation more tractable, we prescribe the following boundary conditions
\begin{equation}
j\left(r_A, t\right)=-\left.D\left(e^{-\beta \mathcal{U}(r)} \frac{\partial}{\partial r}\left(e^{\beta \mathcal{U}(r)} \rho(r, t)\right)\right)\right|_{r=r_A}=0,\label{rbc}
\end{equation}
\begin{equation}
\rho\left(r_m, t\right)=0.\label{abc}
\end{equation}
Applying a reflecting boundary condition at $r_A$ means that the probability flux $j$ at $r_A$ is zero, indicating a reflective boundary where no states flow out of the system. $r_A$ is typically chosen as the initial state for the phase transition dynamics.  Applying an absorbing boundary condition at $r_m$ implies that the probability density $\rho$ at the boundary $r_m$ is zero. This signifies that states are completely absorbed by the boundary, thus the phase transition can be considered complete. And $r_m$ is usually chosen as the point that just crosses the maximum of the thermal potential.

This choice of $r_A$ differs from the one adopted in Ref.~\cite{li2020thermodynamics}. In Ref.~\cite{li2020thermodynamics}, $r_{\mathrm{A}}$ is set to zero for an initial small black hole and to infinity for an initial large black hole. In our study, we focus primarily on the transition dynamics between stable thermodynamic states. Therefore, we set the initial conditions at the local minima of the potential (corresponding to the macroscopic Small Black Hole or Large Black Hole states), as these represent the physical starting points of the phase transition.It is worth noting that even if the initial state were set to $0$ or $\infty$, the system would rapidly evolve toward the nearest stable basin. Consequently, the numerical discrepancy in the MFPT results between these two choices is negligible. In contrast, the Ref.~\cite{li2020thermodynamics} were more concerned with the general evolution from non-equilibrium to equilibrium states; thus, they adopted extreme non-equilibrium configurations as their starting points.

Considering that the probability of finding the state within the interval $(r_A, r_m)$ is given by $\int_{r_A}^{r_m} dr \rho(r, t)$, when the state is not in this interval, we can assume that it has crossed the potential barrier and reached the other state. Therefore, we can regard this integral expression as the probability that the black hole has not undergone the first phase transition by time $t$. This leads directly to the distribution of the MFPT
\begin{equation}
F_p(t)=-\frac{d\left(\int_{r_A}^{r_m} d r \rho\right)}{d t},\label{fpp}
\end{equation}
Based on the defining expression of the MFPT
\begin{equation}
\left\langle t\right\rangle=\int_0^{\infty} d t t F_p(t).\label{<t>}
\end{equation}
the MFPT $\left\langle t\right\rangle$ can be correlated with the probability density $\rho(r,t)$.
By substituting the Fokker-Planck equation (\ref{fp}) and the boundary conditions (\ref{rbc}) and (\ref{abc}) into the expression for $F_p(t)$ (\ref{fpp}) and performing the calculation (\ref{<t>}), we obtain
\begin{equation}
\left\langle t\right\rangle=\frac{1}{D} \int_{r_A}^{r_m} d r \int_{r_A}^r d r^{\prime} e^{\beta\left(\mathcal{U}(r)-\mathcal{U}\left(r^{\prime}\right)\right)}.\label{t}
\end{equation}

In the derivation of the Kramers escape rate, the current $J(r,t)$ is defined as
\begin{equation}
J(r, t)=\frac{D e^{\mathcal{U}\left(r_{\min }\right) / D} \rho\left(r_{\min }, t\right)}{\int_{r_{\min }}^A e^{\mathcal{U}(r) / D} d r},
\end{equation}
where $\mathcal{U}(r)$ is the potential or the generalized free energy in our work, and $D$ is the diffusion coefficient, which can be considered constant when the system reaches thermal equilibrium, and we assume that at $r=A$ (A is any position greater than $r_{\max }$ ), the probability distribution is zero. If we define $p$ as the probability of the state being inside the well or near $r_{\min }$, then we can find
\begin{equation}
p=\rho\left(r_{\min }, t\right) e^{\mathcal{U}\left(r_{\min }\right) / D} \int_{\left(r_{\min }\right)} e^{-\mathcal{U}(r) / D} d r.
\end{equation}

The probability $p$ times the Kramers escape rate $r_k$ is just the current $J(r, t)$, hence we can obtain the escape rate~\cite{risken1989fokker,zwanzig2001nonequilibrium}

\begin{equation}
\frac{1}{r_k}=\frac{p}{J}=\frac{1}{D} \int_{r_{\min }}^A e^{\mathcal{U}(r) / D} d r \int_{\left(r_{\min }\right)} e^{-\mathcal{U}(r) / D} d r.\label{k}
\end{equation}
Physically, the Kramers escape rate represents the fraction of states that successfully surmount the potential barrier per unit time, whereas the MFPT denotes the average time a single state requires to cross the barrier and reach the target state. Consequently, they are approximately reciprocals of each other, as can be readily verified by comparing Eqs.~\eqref{t} and \eqref{k}.

For above two integrals, we can clearly see that the main contribution of the first integral comes from the regions around $r_{\max }$, while the main contribution of the second integral comes from the regions around $r_{\min }$. The Taylor expansions approximation to second order of the potential function $U(r)$ near two extreme points are
\begin{equation}
 \mathcal{U}(r) \approx\mathcal{U}\left(r_{\max }\right)-\frac{1}{2}\left|\mathcal{U}^{\prime \prime}\left(r_{\max }\right)\right|\left(r-r_{\max }\right)^2,
\end{equation}
\begin{equation}
 \mathcal{U}(r) \approx \mathcal{U}\left(r_{\min }\right)+\frac{1}{2} \mathcal{U}^{\prime \prime}\left(r_{\min }\right)\left(r-r_{\min }\right)^2,
\end{equation}
and we may extend the above two integrations boundaries to $\pm \infty$, thus the Kramers escape rate can be taken as~\cite{risken1989fokker,zwanzig2001nonequilibrium}

\begin{equation}
r_k=\frac{\sqrt{\left|\mathcal{U}^{\prime \prime}\left(r_{\min }\right) \mathcal{U}^{\prime \prime}\left(r_{\max }\right)\right|}}{2 \pi} e^{-\frac{\mathcal{U}\left(r_{\max }\right)-\mathcal{U}\left(r_{\min }\right)}{D}} .
\end{equation}

We adopt the exact integral formulation for the MFPT to ensure numerical accuracy across the whole landscape. In contrast, for the Kramers escape rate, we use the approximate analytical form, as it represents a leading-order approximation in the high-barrier limit.

\section{the mean first passage time between black hole phase transition}
\label{4}

\begin{figure}[htb]
	\centering
	\subfigure[$L=15, m_0=10, T=0.017854$.]{
		\includegraphics[width=50 mm]{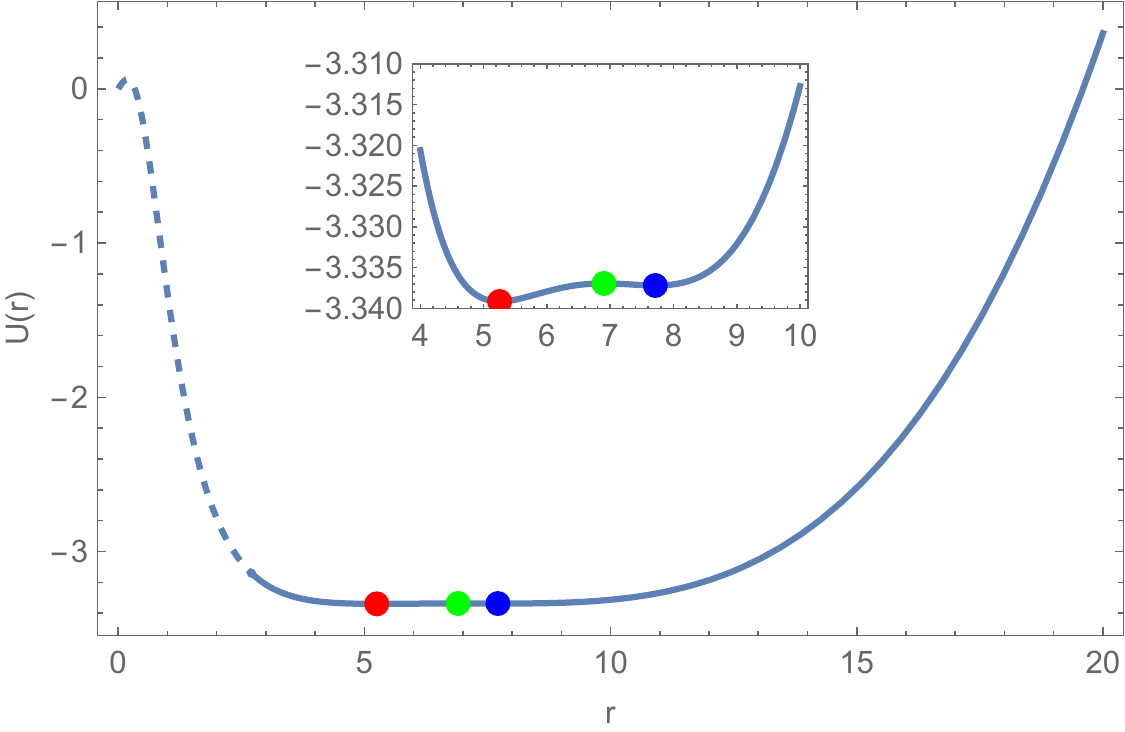} }
	\subfigure[$L=15, m_0=6.5, T=0.018759$.]{
		\includegraphics[width=50 mm]{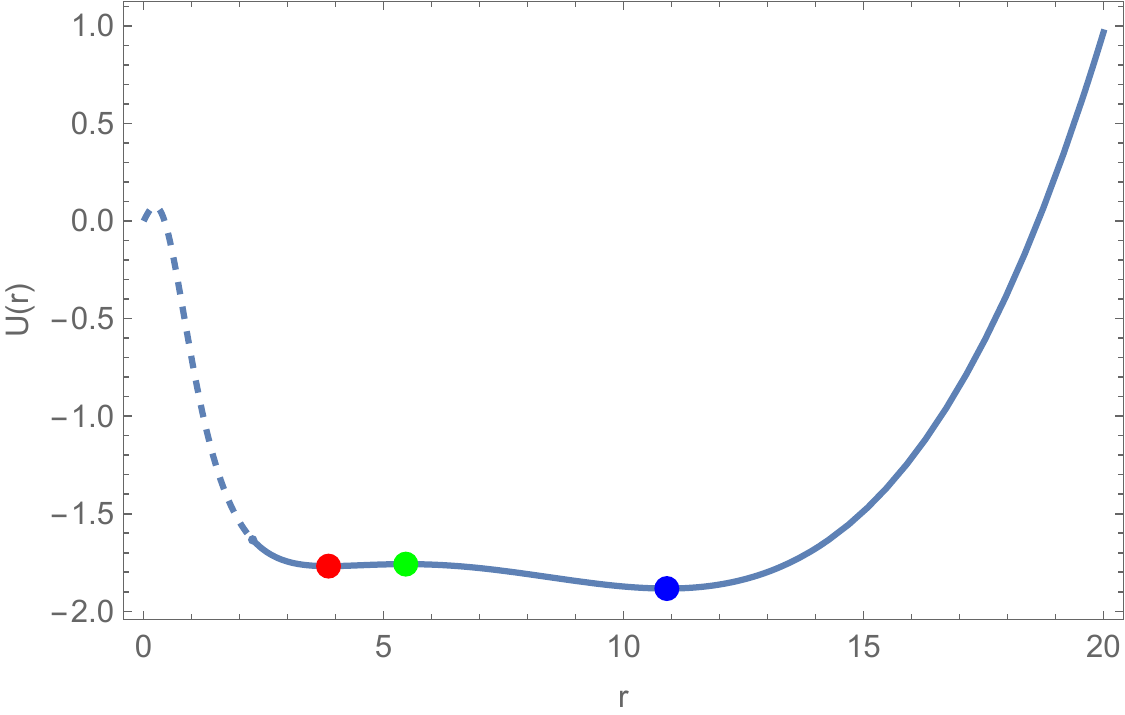} }
	\subfigure[$L=15, m_0=3, T=0.019729$.]{
		\includegraphics[width=50 mm]{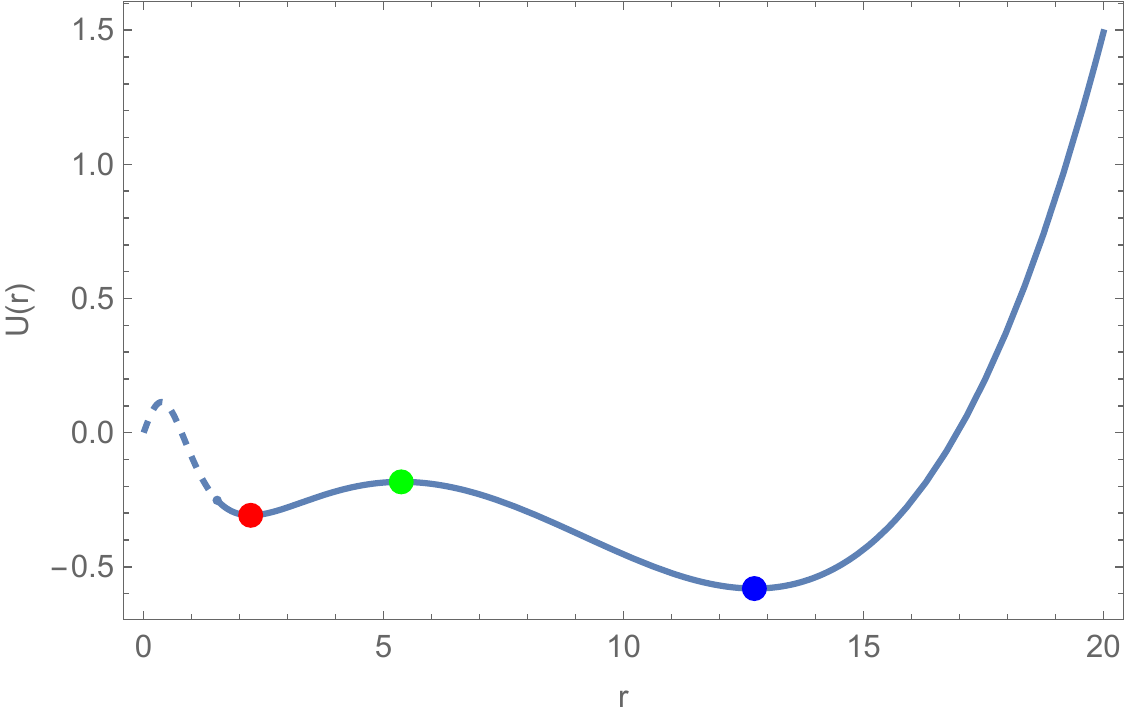} }\\
	\subfigure[$L=10, m_0=2.05, T=0.027750$.]{
		\includegraphics[width=65 mm]{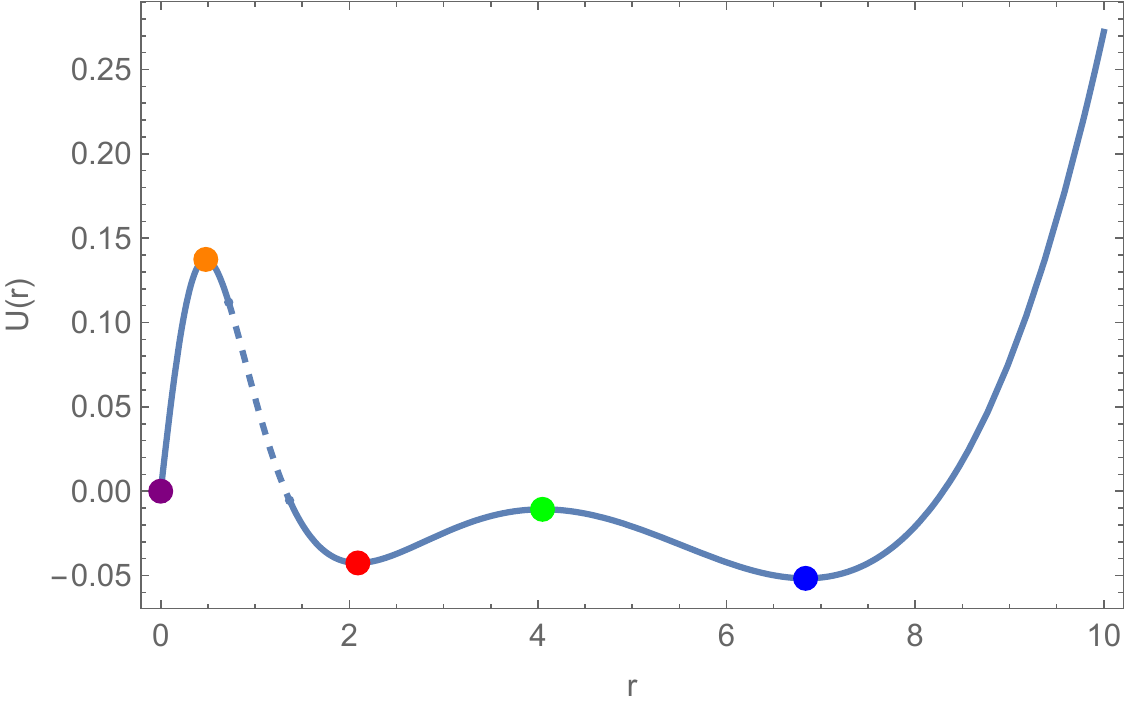} }
	\subfigure[$L=7.67, m_0=1.39, T=0.03550$.]{
		\includegraphics[width=65 mm]{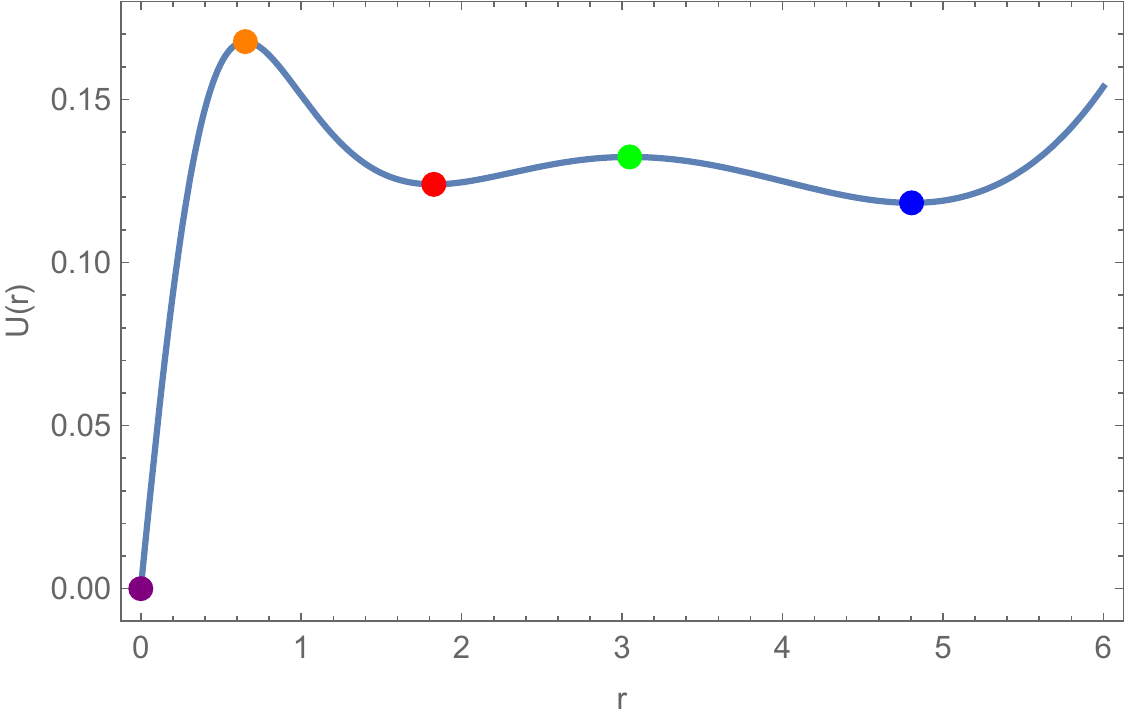} }		
	\caption{The behaviors of generalized free energy. Panels (a)–(c) correspond to Type I black holes, while (d) and (e) represent Type II black holes. The solid lines depict the relationship between the generalized free energy $\mathcal{U}$ and the event horizon $r_+$. In contrast, dashed lines denote non-physical, unattainable regions. The extrema of the generalized free energy are marked with colored dots.}\label{fig2}
\end{figure}

Substituting thermodynamic quantities of the Bardeen-AdS-class black hole into the definition of the generalized free energy (\ref{U}), the generalized free energy $\mathcal{U}(r_+)$ be given
\begin{equation}
\mathcal{U}(r_+)=\frac{r_+}{2}  +\frac{r_+^3}{2 L^2} -\frac{m_0 r_+^3}{2}\left(\frac{q_m}{q_0 (q_0 q_m +r_+^2)} \right)^\frac{3}{2}  -\pi r_+^2 T.
\end{equation}
We set $q_0=1$ to nondimensionalize the thermodynamic quantities. To simplify the ensemble, we set $q_m=q_0$ and varying the parameters $L$ and $m_0$. For convenience, we use $\mathcal{U}(r)$ to denote $\mathcal{U}(r_+)$. At this point, the generalized free energy $U(r)$ simplifies to
\begin{equation}
\mathcal{U}(r)=\frac{1}{2} \left(r +\frac{r^3}{L^2} - m_0 r^3 \sqrt{\left(\frac{1}{1+r^2}\right)^3}  -\pi r^2 T \right).\label{ufree}
\end{equation}
We selected some representative  cases to illustrate the results in the Fig.~\ref{fig2}.
We emphasize once again that $r_+$ cannot always take on a continuous range of values. Since the order parameter $r$ (or $r_+$) must correspond to a physically existing black hole when constructing the generalized free energy landscape, the condition $T_h(r) \geq 0$ must be strictly satisfied. Accordingly, the regions that do not fulfill this requirement are represented by dashed lines in Fig.~\ref{fig2}.
In fact, these dashed lines represent the generalized free energy calculated using the ``temperature'' derived from the inner horizon radius. Consequently, their thermodynamic behavior is physically meaningless.

It is readily apparent that the behavior of the generalized free energy is classified into three cases. 
Case 1 is defined only in the latter half of the curve and corresponds to the Type I black holes in~\cite{wu2025thermodynamics}. Case 2 is undefined in some intermediate region. Case 3 is definable across the entire interval. Both case 2 and case 3 curves correspond to the Type II black holes discussed in reference~\cite{wu2025thermodynamics}. Since we are primarily concerned with phase transitions that feature an intermediate state of a regular black hole, and Type II black holes are unable to form a regular black hole, we will focus our analysis on the case 1 curves. 

It should be noted that as certain regions of the Case 2 curves are not well-defined, our current approach is unable to analyze the phase transition behavior from the vacuum state to the black hole state in this specific case. Consequently, we must rely on traditional black hole thermodynamic analysis for these scenarios; for a detailed discussion, please refer to the work of Ref.~\cite{wu2025thermodynamics}.

We found that the generalized free energy curves in Fig.~\ref{fig2}(a)-(c) are very similar, yet they exhibit differences. If we calculate the event horizon radius ($r_+$) for the regular black hole in these three cases—specifically, the Bardeen-AdS-class black hole—we observe the following: The event horizon radius in Fig.~\ref{fig2}(a) is located at the potential well corresponding to the Small Black Hole state. The event horizon radius in Fig.~\ref{fig2}(b) is located at the potential barrier corresponding to the intermediate unstable black hole state. The event horizon radius in Fig.~\ref{fig2}(c) is located at the potential well corresponding to the Large Black Hole state.

\begin{figure}[htb]
	\centering
	\subfigure[$L=15, m_0=10$.]{
		\includegraphics[width=50 mm]{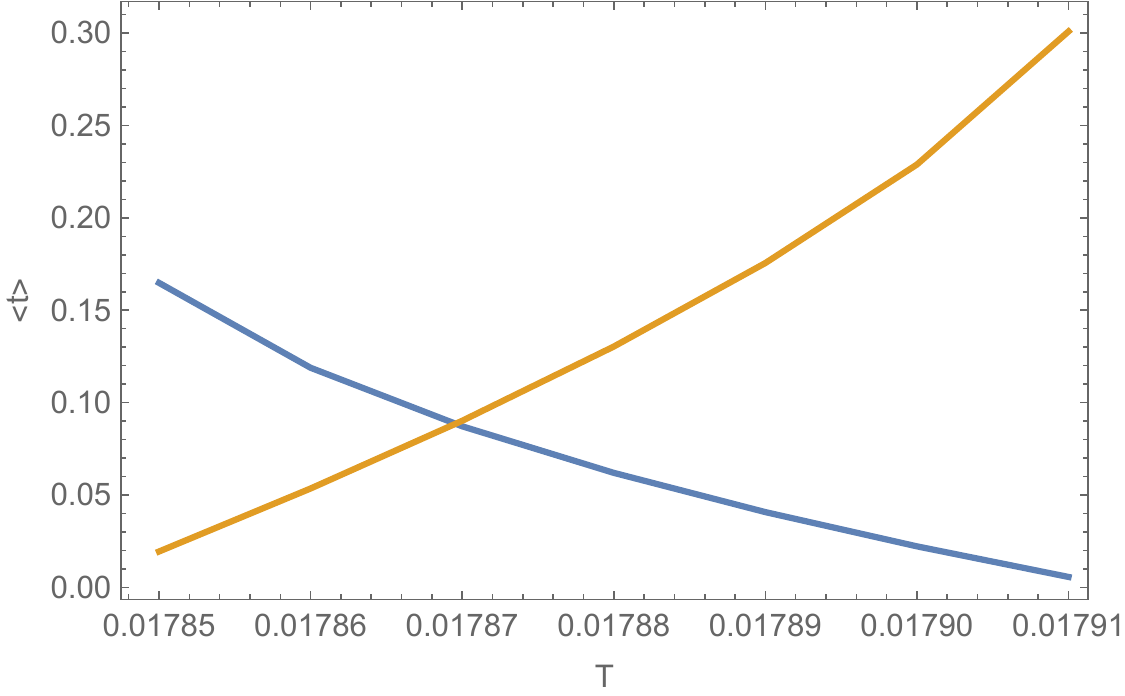} }
	\subfigure[$L=15, m_0=6.5$.]{
		\includegraphics[width=50 mm]{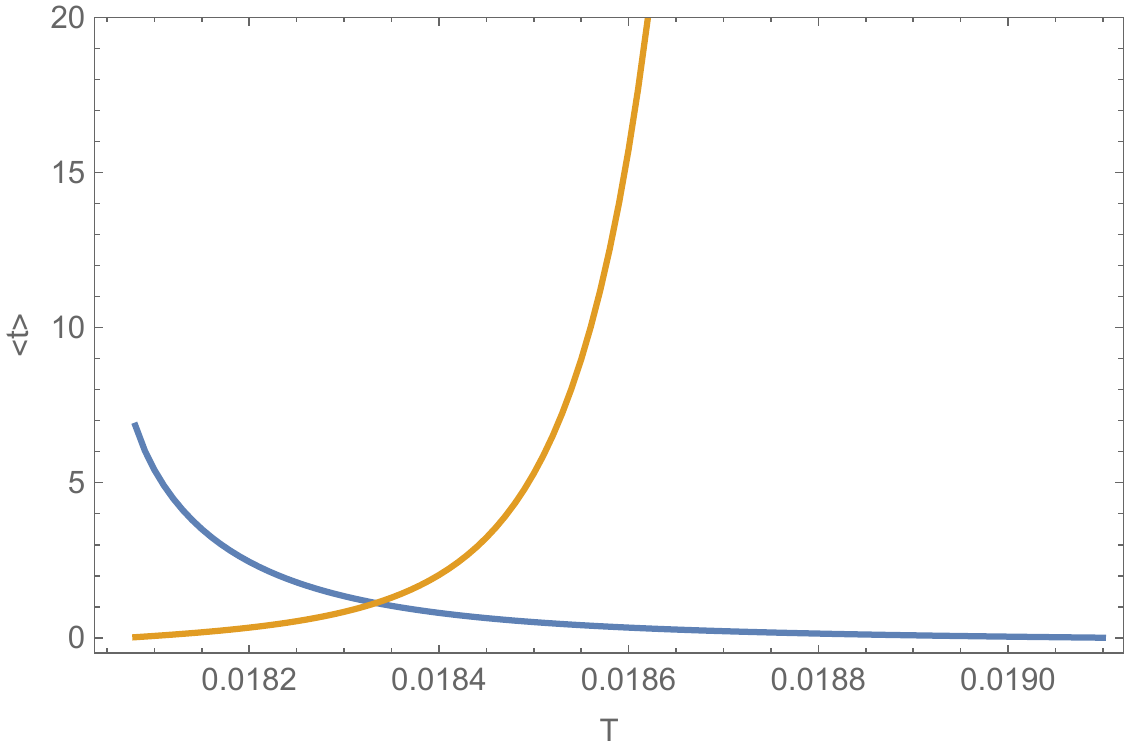} }
	\subfigure[$L=15, m_0=3$.]{
		\includegraphics[width=50 mm]{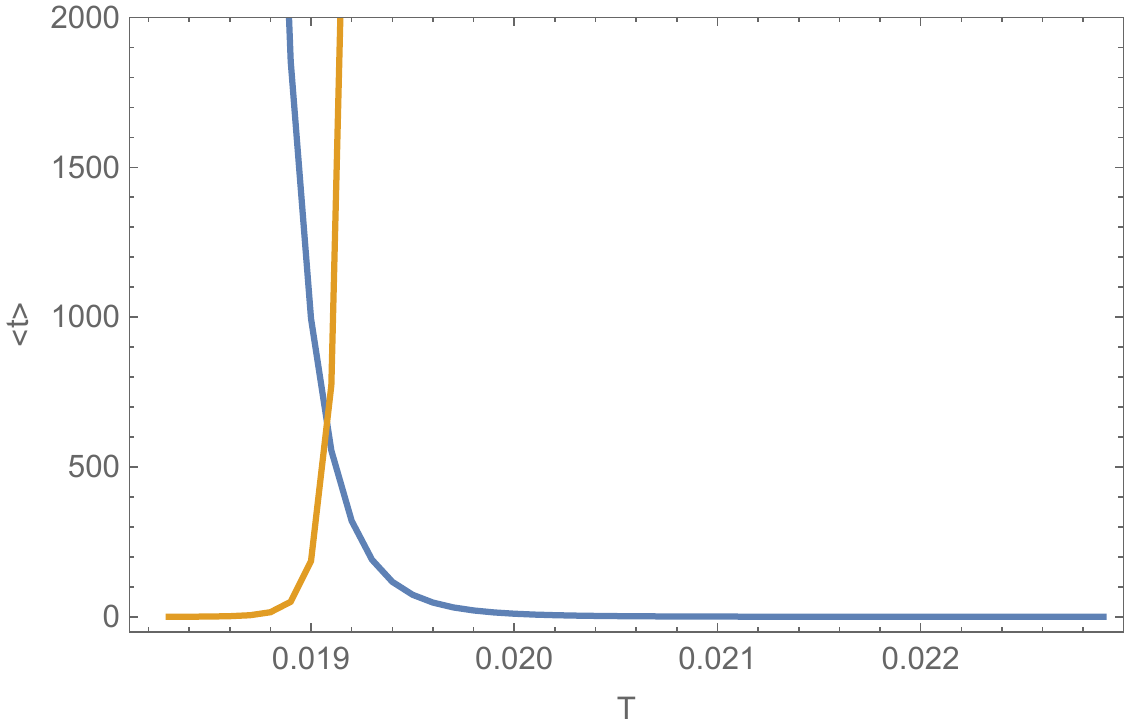} }
	\caption{The mean first passage time with respect to the temperature for type I black hole. The blue curve represents the transition process from the small black hole state to the large black hole state, while the orange curve depicts the reverse process from the large black hole state to the small black hole state; the two curves intersect at a single point.}\label{fig3}
\end{figure}

\begin{figure}[htb]
	\centering
	\subfigure[$L=15, m_0=10$.]{
		\includegraphics[width=50 mm]{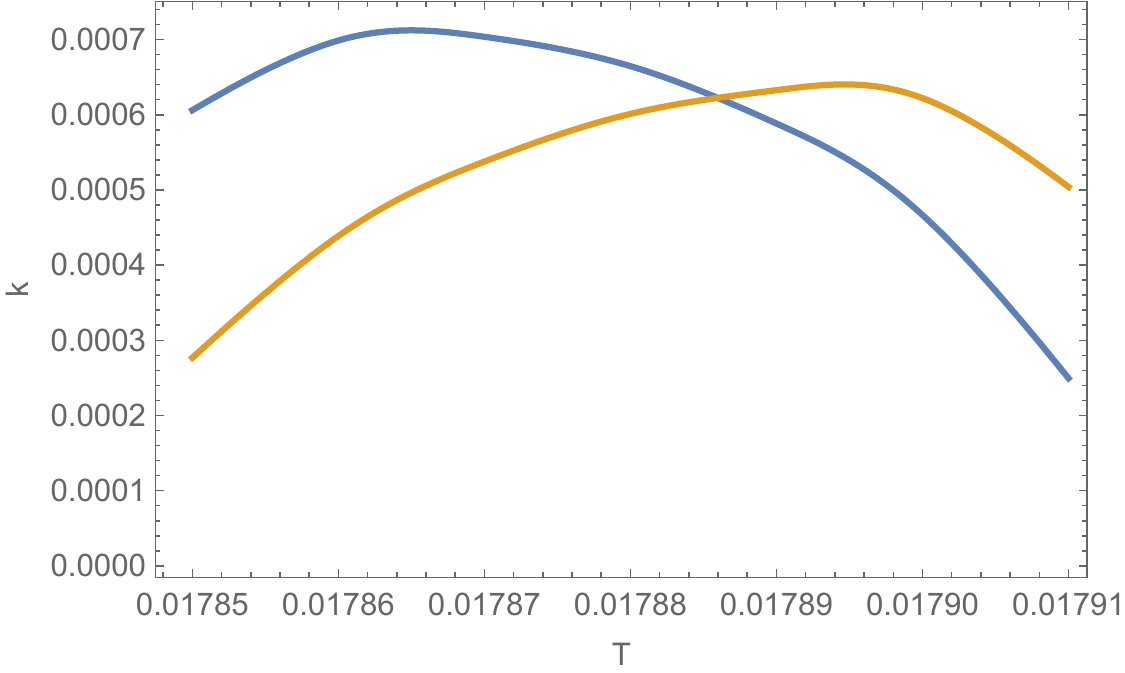} }
	\subfigure[$L=15, m_0=6.5$.]{
		\includegraphics[width=50 mm]{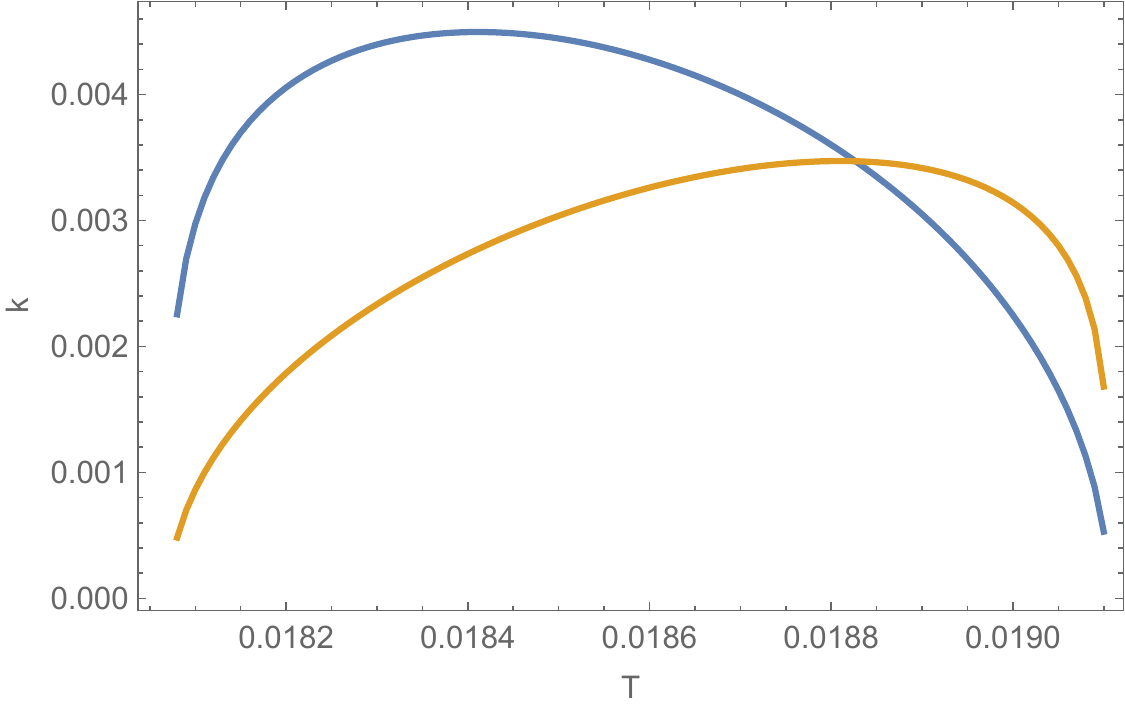} }
	\subfigure[$L=15, m_0=3$.]{
		\includegraphics[width=50 mm]{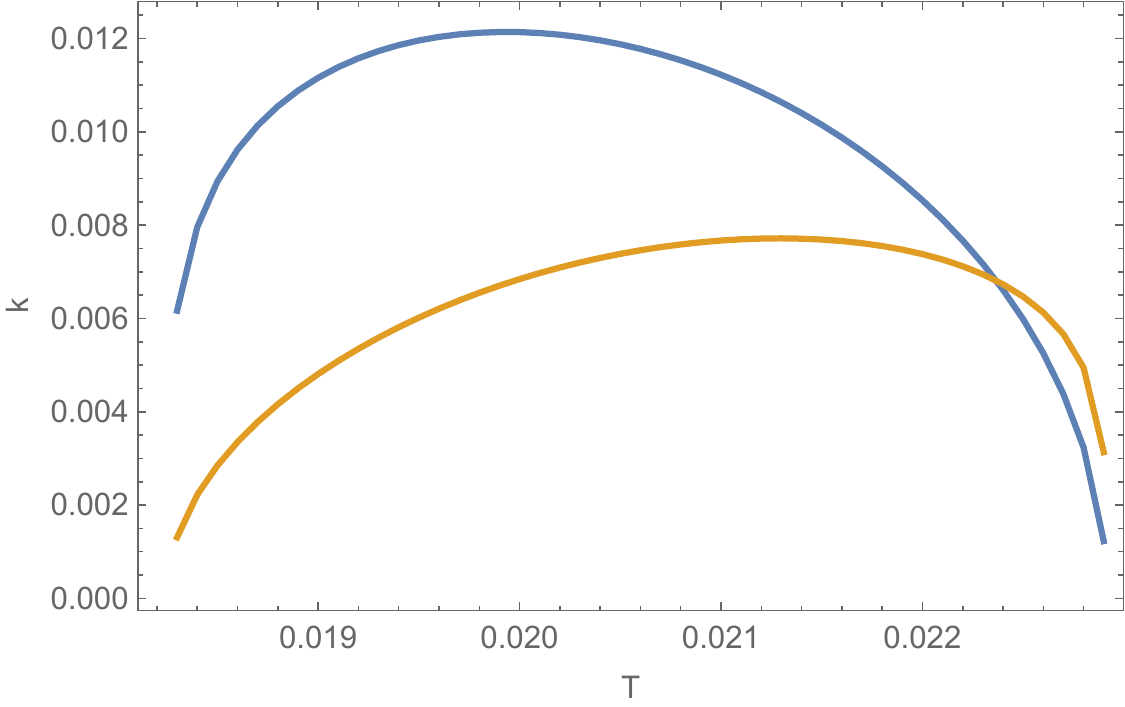} }
	\caption{The Kramers escape rate with respect to the temperature for type I black hole. The blue curve represents the transition process from the small black hole state to the large black hole state, while the orange curve depicts the reverse process from the large black hole state to the small black hole state; the two curves intersect at a single point.}\label{fig5}
\end{figure}

We analyze the phase transition between small and large black holes by calculating the MFPT and the Kramers escape rate. The results of this analysis are plotted in Fig.~\ref{fig3} and Fig.~\ref{fig5}. The blue curve represents the process from the small black hole state to the large black hole state, while the orange curve represents the process from the large black hole state to the small black hole state. As shown in Fig.~\ref{fig3}, in all three cases, the MFPT for the small-to-large black hole state transition gradually decreases as the temperature $T$ increases, and the MFPT for the large-to-small black hole state transition gradually increases. The intersection point of the two curves represents the dynamic equilibrium where the forward and reverse processes balance, and their MFPTs are exactly equal. 
The behavior of the Kramers escape rate is consistent across the three cases, all exhibiting an initial increase followed by a decrease. At lower temperatures, the escape rate from small to large black holes is higher, while at higher temperatures, the opposite situation occurs. The intersection point represents the state of dynamic equilibrium.

However, the three cases differ in the presence of a regular black hole state: In Fig.~\ref{fig3}(a) (or Fig.~\ref{fig5}(a)), the regular black hole state exists at the small black hole position. In Figure 3-c, the regular black hole state exists at the large black hole position. In both of these cases (Fig.~\ref{fig3}(a)(c) or Fig.~\ref{fig5}(a)(c)), the phase transition process involves a stable (or metastable) regular black hole state.  In Fig.~\ref{fig3}(b) (or Fig.~\ref{fig5}(b)), the regular black hole state is situated at the unstable potential barrier between the small and large black hole states. Consequently, the phase transition process in this case traverses an unstable regular black hole state.

\begin{figure}[htb]
	\centering
	\subfigure[]{
		\includegraphics[width=65 mm]{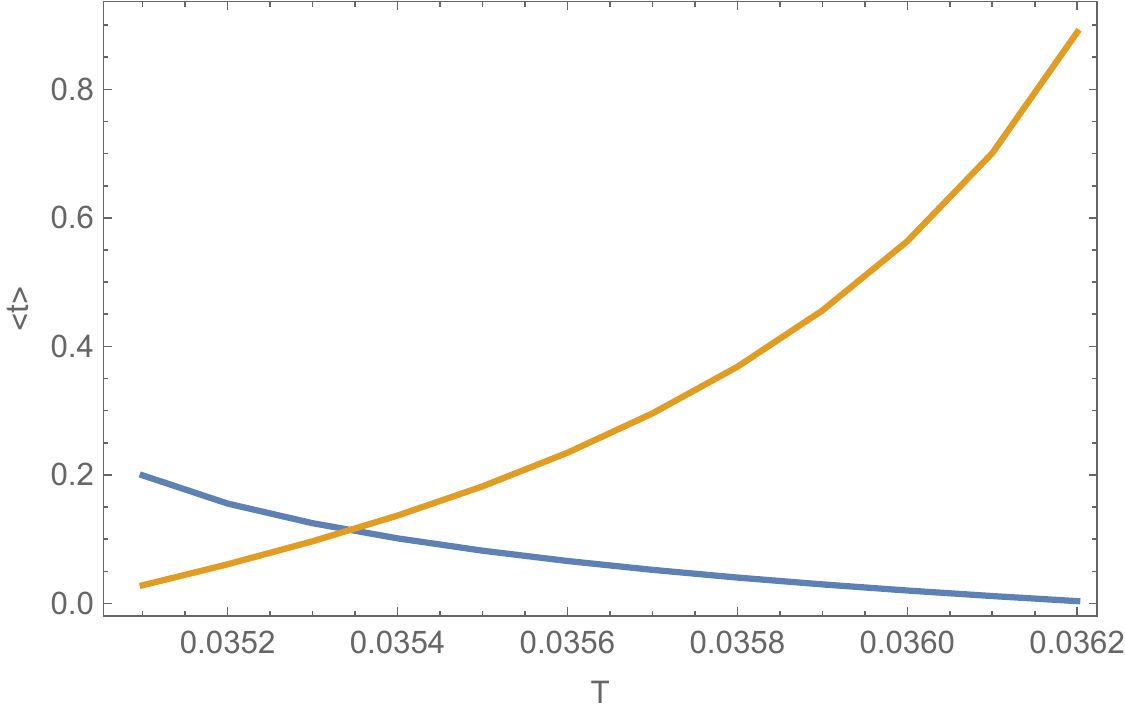} }
	\subfigure[]{
		\includegraphics[width=65 mm]{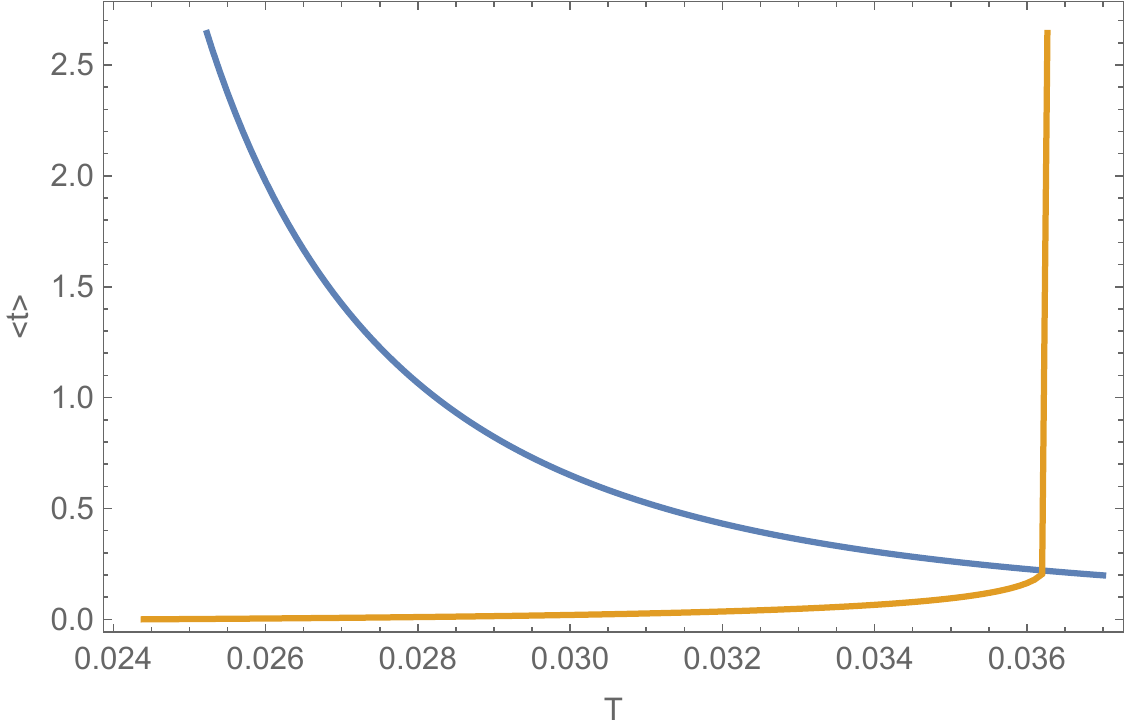} }
	\caption{The mean first passage time with respect to the temperature for type II black hole. (a)The blue curve represents the transition process from the small black hole state to the large black hole state, while the orange curve depicts the reverse process. The two curves intersect at a single point. (b)The blue curve represents the transition process from the vacuum state to the black hole state, while the orange curve depicts the reverse process.  The orange curve exhibits a mutation point.}\label{fig4}
\end{figure}

\begin{figure}[htb]
	\subfigure[]{
		\includegraphics[width=50 mm]{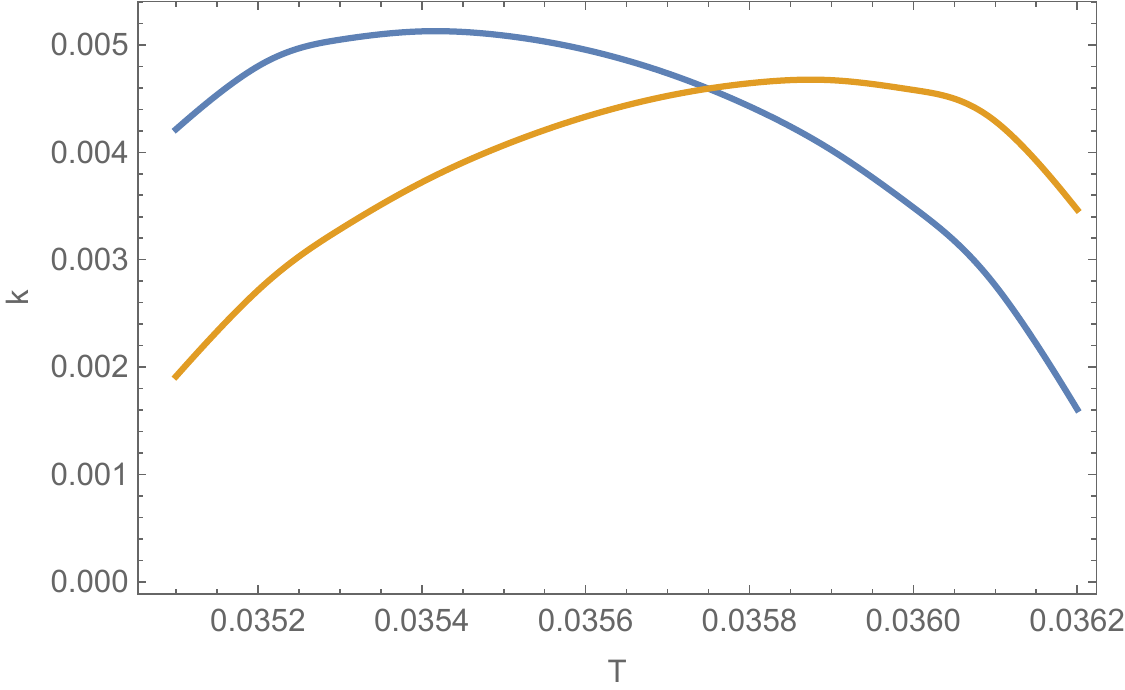} }
	\subfigure[]{
		\includegraphics[width=50 mm]{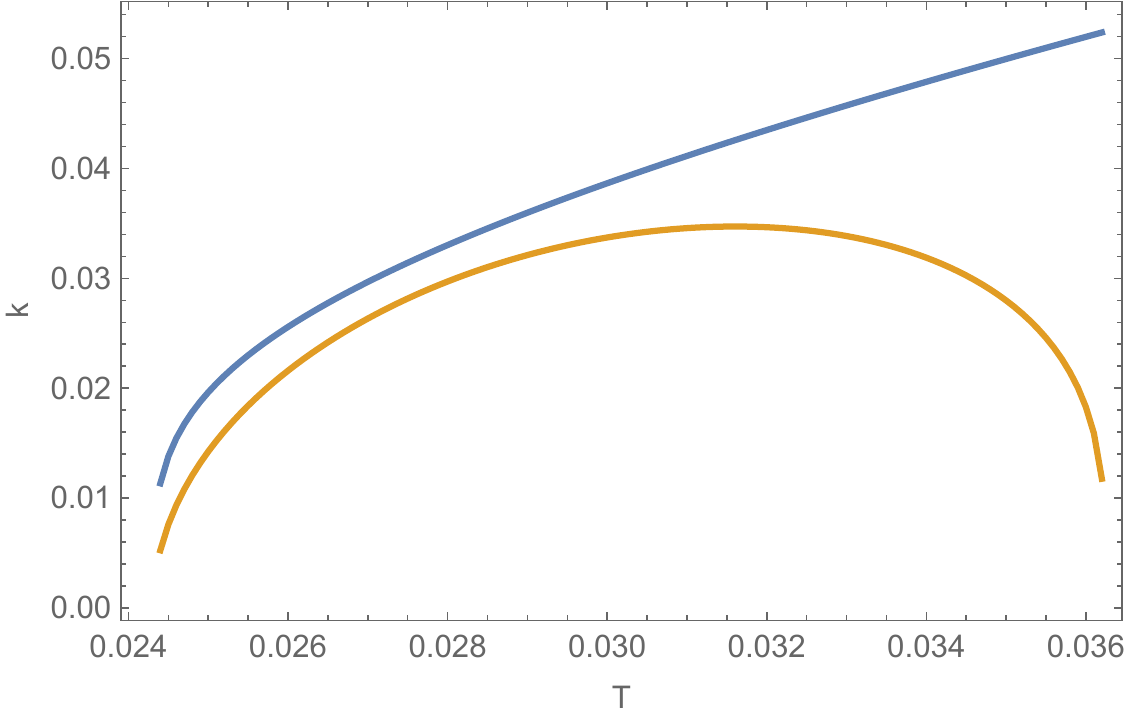} }
	\subfigure[]{
		\includegraphics[width=50 mm]{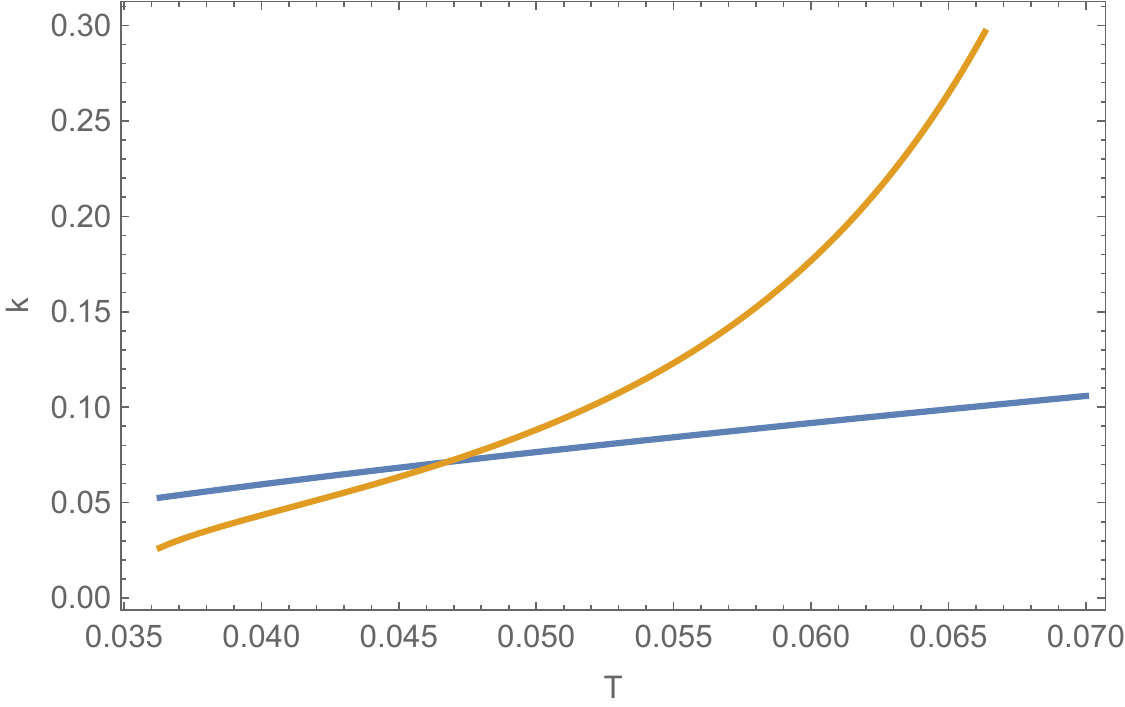} }
	\caption{The Kramers escape rate with respect to the temperature for type II black hole. (a)The blue curve represents the transition process from the small black hole state to the large black hole state, while the orange curve depicts the reverse process. (b)The blue curve represents the transition process from the vacuum state to the large black hole state, while the orange curve depicts the reverse process. (c)The blue curve represents the transition process from the vacuum state to the small black hole state, while the orange curve depicts the reverse process.}\label{fig6}
\end{figure}

Additionally, we also investigated the phase transition between Type II black holes and the vacuum state. The resulting data is plotted in Fig.~\ref{fig4} and Fig.~\ref{fig6}. It is worth noting that regular black holes are absent in this case.

At low temperatures, the system consists of a large black hole state and a vacuum state. As shown in Fig.~\ref{fig4}(b) (or Fig.~\ref{fig6}(b)), the curves for the MFPT and the Kramers escape rate between the vacuum and the large black hole do not intersect, indicating that no phase transition occurs between them. As the temperature $T$ increases, the small black hole state emerges. At this point, the dynamic process between the vacuum and the large black hole terminates, and the small black hole begins to participate in the stochastic dynamics. According to Fig.~\ref{fig4}(a) (or Fig.~\ref{fig6}(a)), as $T$ continues to rise, the MFPT from the small black hole to the large black hole gradually decreases, while the MFPT from the large black hole to the small black hole increases; their intersection signifies a state of dynamic equilibrium. A similar process occurs between the vacuum and the small black hole, as illustrated in Fig.~\ref{fig4}(b) (or Fig.~\ref{fig6}(c)). However, because the transition from the small black hole to the vacuum state occurs more readily when the small black hole and the large black hole reach dynamic equilibrium, the equilibrium between the small black hole and the large black hole does not constitute a phase transition. Consequently, the only authentic phase transition in the entire process exists between the vacuum state and the small black hole state.

Theoretically, the Kramers rate and the MFPT should be inversely proportional in the absence of calculation errors. However, this relationship does not strictly hold in our numerical results for the following reasons. While we employ an exact integral formula for the MFPT, the Kramers rate is derived using a second-order Taylor expansion at the local maxima and minima of the potential $\mathcal{U}$ to obtain an approximate algebraic expression. Consequently, the inverse proportionality holds only when the potential wells and barriers deviate minimally from a quadratic function. In general,  this requires the thermal potential to satisfy the deep well condition, namely $\Delta \mathcal{U} =\mathcal{U}\left(r_{\max }\right)-\mathcal{U}\left(r_{\min }\right) \gg D$. In this regime, even if the potential is globally non-linear, the integral remains sensitive only to the second-order profile near the extrema due to the rapid decay of the exponential factor $e^{-\Delta \mathcal{U}/D}$. However, as shown in Fig.~\ref{fig2}, our chosen potential $\mathcal{U}$ does not satisfy this condition, leading to the observed discrepancy.      Nevertheless, since the Kramers rate is purely algebraic, it is significantly faster to compute than the MFPT, making it a convenient tool for qualitative discussions. For precise quantitative calculations, the MFPT method remains indispensable. These discussions have been incorporated into the revised manuscript.

\section{Summary}
\label{5}

We acknowledge the established challenge, the energy and magnetic charge of regular black holes are generally correlated with the matter field's coupling coefficients. Consequently, their associated integration constants cannot be decoupled from these coupling parameters. In thermodynamics, this necessitates that the Fig.~\ref{fig3} and Fig.~\ref{fig5} variation of energy and other coupling constants be considered simultaneously, which leads to an inconsistency between the thermodynamic quantities and the First Law.
To circumvent this, the integration constants (such as energy and charge) are treated not as coupling parameters but as independent variables, thereby exploring a more generalized black hole solution.
Using the generalized free energy landscape, we determine the thermal potential. Based on this potential, we analyzed the MFPT and the Kramers escape rate to quantify the dynamics of the black hole phase transition.

For Type I black holes, the results indicate that the large black hole state is more stable at lower temperatures, while the small black hole state becomes more stable at higher temperatures. Since a dynamic equilibrium exists between them, a phase transition from the large black hole to the small black hole occurs as $T$ increases.
However, the presence of the regular black hole gives rise to three distinct scenarios, depending on whether it identifies with the large black hole state, the small black hole state, or the intermediate unstable state. In the first two cases, the black hole phase transition process passes through a stable (or metastable) regular black hole state. In the third case, the phase transition process involves an unstable regular black hole state as an intermediate stage.

For Type II black holes, the results indicate that at low temperatures $T$, the system comprises a large black hole  state and a vacuum state, with the latter being more stable during their dynamic interaction. As $T$ increases, a small black hole state emerges, causing the transition between the vacuum and large black hole states to cease; instead, the vacuum state begins to interact dynamically with the small black hole state. Meanwhile, a dynamic process also exists between the small black hole and large black hole states. Upon the emergence of the small black hole state, the large black hole state is more stable than the small black hole state but less so than the vacuum state. As $T$ continues to increase, the stability of the small black hole state first surpasses that of the large black hole state and subsequently exceeds that of the vacuum state. In other words, the vacuum state becomes the most stable at very high temperatures. Consequently, despite the dynamic equilibrium between small black hole and large black hole states, this process does not constitute a phase transition; rather, the phase transition occurs exclusively between the vacuum and small black hole states.

This detailed process description provides an overall picture of the thermodynamic phase transition of the Bardeen-AdS-class black hole, which deepens our understanding of the stochastic thermodynamic behavior of black holes. Moreover, this research approach can be extended to higher order gravity models, which exhibit a rich variety of complex phase transition behaviors, thereby enabling us to obtain dynamic information on the thermodynamic phase transitions of black holes.

\section*{Acknowledgments}
We sincerely thank the anonymous reviewers for their insightful suggestions, which have significantly enhanced this work. This research was supported in part by the National Natural Science Foundation of China (Grant Nos. 12275216, 12575064, 12247103), by the Natural Science Basic Research Plan in Shaanxi Province of China (Grant No. 2025JC-YBQN-029).

\end{document}